\documentclass[amsmath,amssymb,ams, preprint, superscriptaddress,nofootinbib, twocolumn]{article}
\usepackage[utf8]{inputenc}
\usepackage{lineno}
\usepackage[a4paper,textwidth=17.9cm,textheight=26cm]{geometry}
\usepackage[none]{hyphenat}
\usepackage[english]{babel}
\usepackage[explicit]{titlesec}
\usepackage[T1]{fontenc}
\usepackage{lmodern}
\usepackage{helvet}
\usepackage{lineno}
\usepackage{graphicx}
\usepackage{xr}
\usepackage{authblk}
\usepackage{upgreek}
\usepackage{caption}
\usepackage{amsmath}
\usepackage{chemformula}
\usepackage{dblfloatfix} 
\usepackage{xcolor}
\usepackage{bm}
\usepackage{float}
\usepackage{cite}

\begin{document}
\title{Tailoring Transverse Magneto-Optical Kerr Effect Enhancement in Mie-resonant Nanowire-based Metasurfaces}
\author{Karen A. Mamian} \date{}
\author{Vladimir V. Popov} \date{}
\author{Aleksandr Yu. Frolov} \date{}
\author{Andrey A. Fedyanin} \date{}
\affil{\textit{Faculty of Physics, Lomonosov Moscow State University, Moscow 119991, Russia}}

\maketitle

\begin{abstract}

Enhancement and tailoring of the transverse magneto-optical Kerr effect (TMOKE) in hybrid metasurfaces comprising rectangular silicon nanowires coupled with a nickel substrate are demonstrated. The excitation of Mie modes of different orders in nanowires causes the enhancement. The in-plane magnetic dipole mode leads to the largest TMOKE enhancement compared to other Mie modes. Changing the width of silicon nanowires leads to a modification of that mode, thereby ensuring the tailoring of the TMOKE within the range of 2.2\% -- 3.8\%. This tunability is associated with the modification of the near-field localized at the Si/Ni interface and the far-field response of the excited magnetic dipole mode. Adjusting these two quantities allows one to achieve the highest values of the TMOKE caused by individual Mie modes in silicon nanowires.

\end{abstract}

\section{Introduction}

Efficiently controlling the light properties at the nanoscale level 
is crucial for modern nanophotonics. A promising way to achieve this control is to exploit the magneto-optical effects, which occur when light interacts with the magnetized media. These effects can be used to control light intensity, phase, and polarization through the use of an external magnetic field in a variety of applications including magneto-optical light modulators~\cite{optmodul}, sensors~\cite{Belyaev2020, Murzin2023, Grunin2016, Wang2021, Takashima2020}, data writing devices~\cite{magn_recording}, Kerr microscopes~\cite{kerr_micr, McCord2015}, and others. 

The magneto-optical effects in the most common magneto-optically active materials, such as ferromagnetic metals, do not typically reach the levels required for applications~\cite{Krinchik1968MagnetoopticalPO}. Advances in fabrication technologies have enabled the design of resonant nanostructures. Notably, magneto-optical effects can now be enhanced and tuned by tailoring the optical resonances due to a high localization of electromagnetic fields in subwavelength volumes~\cite{Maccaferri2023, Salazar2022_a, 10.1063/5.0072884}. The early enhancement methods were based on the use of ferromagnetic thin films~\cite{Ferguson}, magnetoplasmonic crystals~\cite{Grunin, Temnov2018, Belotelov_2, Kiryanov2022, Frolov2020, Dyakov2019, comb_noble_magn, MacCaferri2015,Chetvertukhin2012, Bonanni2011-wm, Enhanced2023, Takashima2022, Kataja2015, Chekhov:14, Makarova2024}, and magnetic dielectric materials with noble metal plasmonic nanoparticles~\cite{Wu2008, CHETVERTUKHIN2015110, Musorin2019}. However, the enhancement in these cases is limited by the ohmic losses in the metal. 

\begin{figure*}[!b]
	\centering
	\includegraphics[width=1\linewidth]{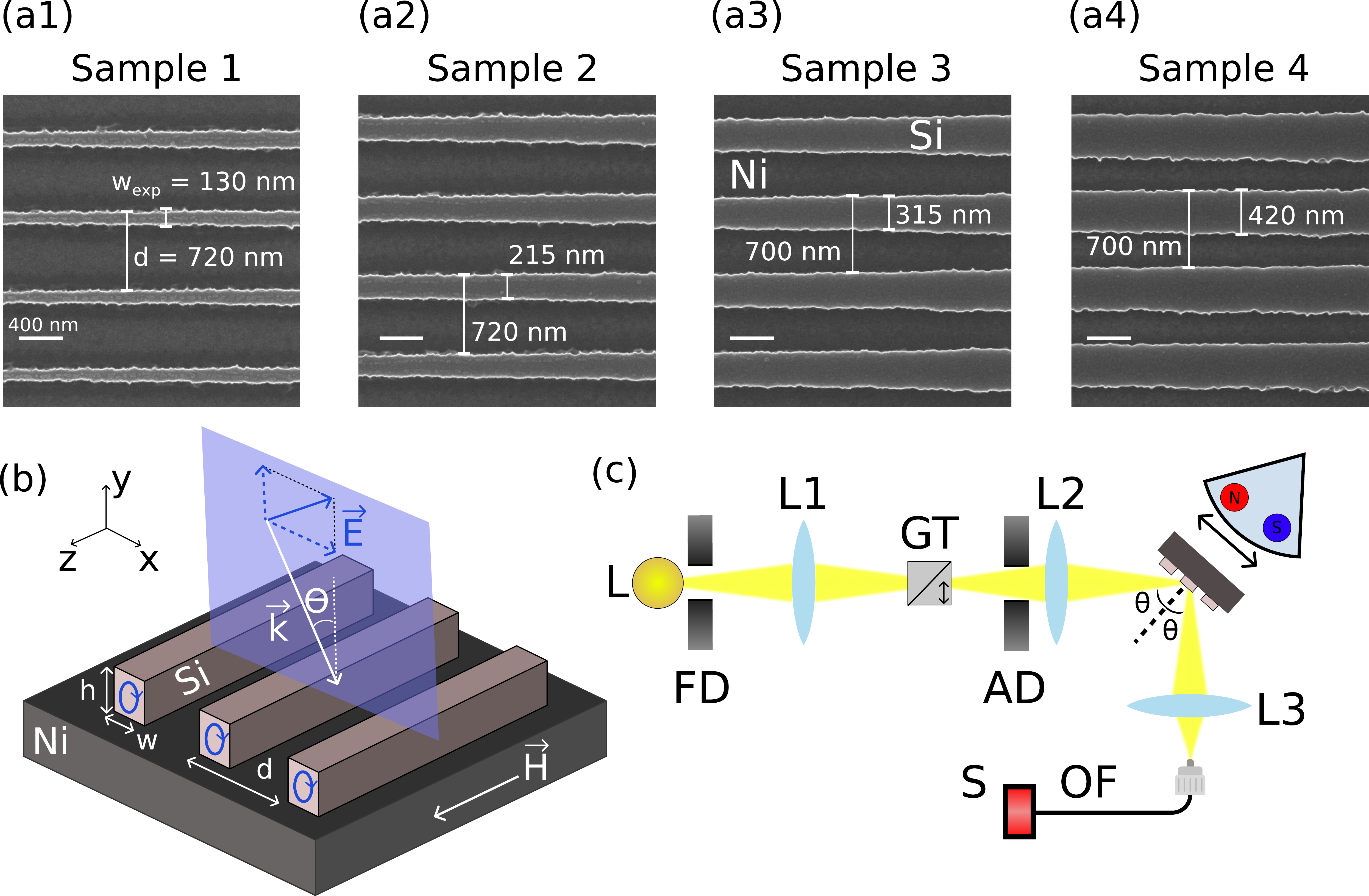}
	\caption{(a1 -- a4) SEM images of the four arrays of silicon nanowires with different widths ($w$) placed on the nickel substrate. (b) Sample scheme and incident light configuration. (c) Experimental setup: L --- lamp; FD --- field diaphragm; AD --- aperture diaphragm; L1, L2, L3 --- lenses; GT --- Glan-Taylor prism; OF --- optical fiber; S --- spectrometer.}
	\label{fig:0}
\end{figure*}

Magnetophotonic crystals~\cite{mphotcrystal, Inoue_2006} were the first dielectric resonant magneto-optical nanostructures proposed to overcome these losses. Further developments in nanophotonics focus on leveraging the properties of high refractive index dielectric metasurfaces~\cite{Bohren2008absorption, Brongersma1, Zhang2018}. They support electric and magnetic Mie resonances with high field localization, geometrical tunability of the resonant wavelength, and low absorption losses~\cite{lowloss, Evlyukhin2010, Garcia, Fu2013, Demonstration_of_Magnetic_Dipole_Resonances, Kuznetsov2012}. Mie modes are induced by circular displacement currents, and their near-field distributions are confined inside the dielectric. Thus, two possible scenarios of magneto-optical enhancement emerge. On the one hand, magnetic dielectric-based metasurfaces can be used to localize light directly within the magnetic material~\cite{Ignatyeva2020}. However, this approach suffers from a relatively low refractive index and gyration constant of typical magnetic dielectric materials (i.e. Bi~-~substituted iron garnet). On the other hand, one could use high-index non-magnetic dielectric resonant nanostructures to localize light near the magnetic material, leveraging the fields at the interface. The previously studied disk-based metasurfaces covered with or placed on ferromagnetic~\cite{nanolabMiemagnetic, Barsukova2019} and magnetic dielectric materials~\cite{Krichevsky2021, arxiv_belotelov} indeed provide a large magneto-optical response. However, in this approach it is crucial to properly optimize the interaction area between the magnetic material and Mie resonant nanostructures. 
Our idea is to use one-dimensional arrays of long rectangular nanowires that support the same Mie modes as in the case of nanodisks. In such a nanowire geometry, the Mie mode field distribution is more sensitive to the external magnetic field because of a more homogeneous field distribution along one of the dimensions~\cite{Wiecha2017, Ee2015, Landreman2016, RNW, Gholipour2017, Karvounis2019, Cao2009, Mamian2024} than can be observed in the nanodisk case. This allows for both a stronger TMOKE enhancement and tailoring in magnitude compared to previous works.

In this paper, we demonstrate a significant TMOKE enhancement in hybrid Si-Ni metasurfaces based on a one-dimensional Mie-resonant rectangular nanowire array. The magnitude, as well as the spectral features of TMOKE and the figure of merit, are controlled by varying the width of silicon nanowires. The observed tunability is associated with a modification of the magnetic dipole mode that causes a change in the near-field, localized in the nickel layer, and the far-field response of the structure. 

\section{Samples and methods}
Our samples are hybrid metasurfaces formed by rectangular amorphous silicon nanowires periodically arranged on a nickel substrate, which are fabricated by a combination of electron-beam and lift-off lithography techniques. The parameters were designed to ensure the excitation of Mie resonances in the nanowires in the visible and near-infrared spectral ranges. The width of the nanowires is the main parameter that varies across the samples, while their period and height are kept relatively constant. Scanning electron microscopy (SEM) and atomic force microscopy (AFM) were utilized to characterize the samples. Indeed, the samples have approximately the same height, varying in the range from $170$~nm to $180 $~nm as shown in the AFM images in Figure~S1 of Supporting materials. SEM images in Figure~\ref{fig:0}~(a1 -- a4) demonstrate the width $w$ and the period $d$ of the nanowires: $w~=~130$~nm, $d = 720 $~nm; $w=215$~nm, $d = 720 $~nm; $w=315$~nm, $d = 700 $~nm; $w=420$~nm, $d = 700 $~nm for Samples 1, 2, 3, and 4, respectively. Thus, the period of the nanowires is also approximately constant across the samples. Note that the $d$ and $w$ values change along the nanowire axis.

\begin{figure*}[!b]
	\includegraphics[width=1\linewidth]{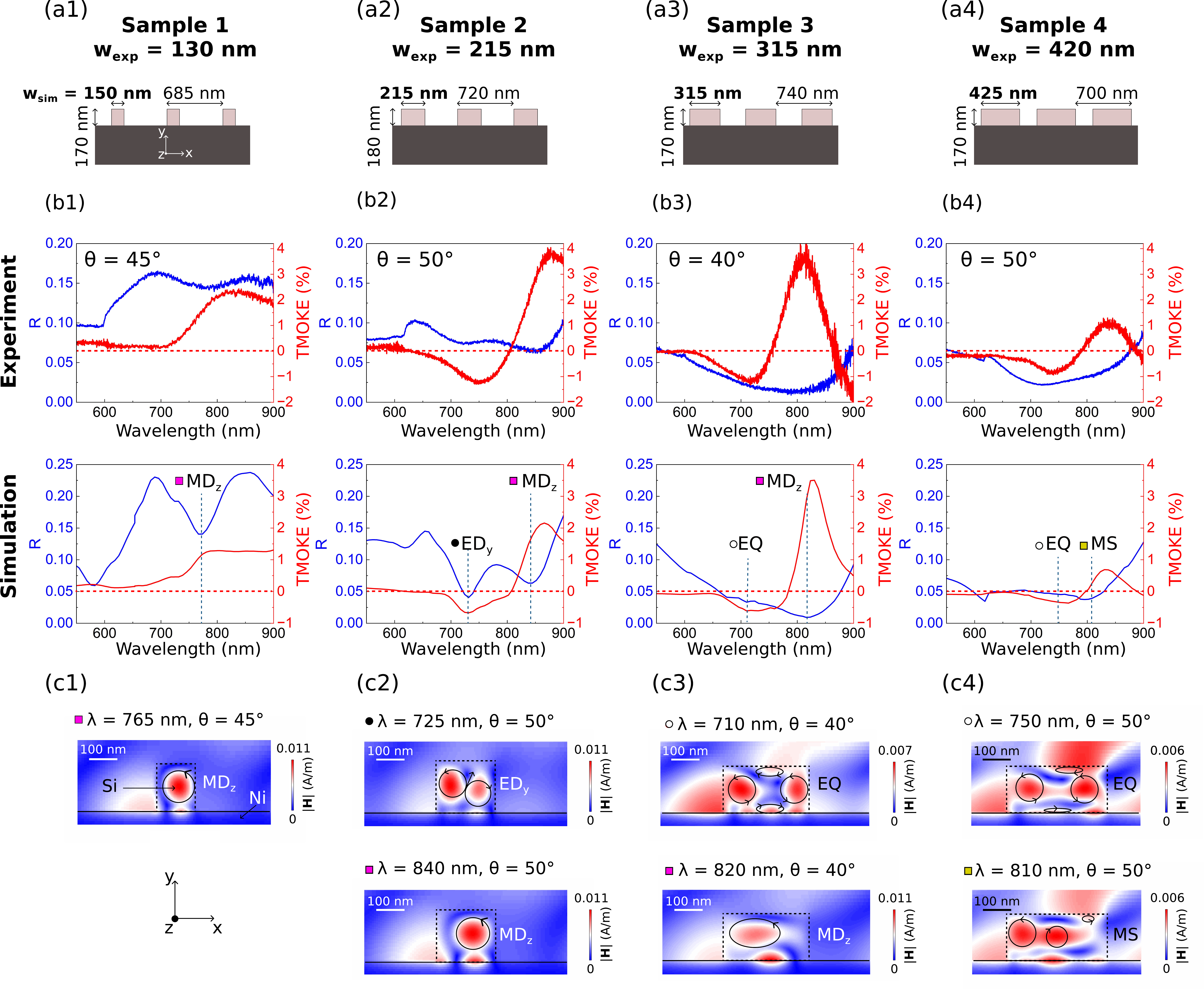}
	\caption{(a1 -- a4) Schematic cross-sections of the Si-Ni metasurfaces and geometrical
parameters used in simulations. (b1 -- b4) Experimental (first row) and simulated (second row) TMOKE (red curves) and reflectance (blue curves) spectra for Samples 1 -- 4 at optimal angles of light incidence. (c1 -- c4) Simulated distributions of $|\textit{\textbf{H}}(x,y)|$ across the silicon nanowire and nickel substrate. The black dashed curves show the nanowire border, and the thick horizontal lines represent the nickel surface. The direction of the electric field of modes is schematically shown by black arrows. MS stands for mode superposition.}
	\label{fig:1}
\end{figure*}
The schematics for the reflectance and TMOKE measurements in hybrid Si-Ni metasurfaces are shown in Figure~\ref{fig:0}~(b) and (c), respectively. The light from a halogen lamp passed through two lenses (L1 and L2), field and aperture diaphragms (FD and AD), and a Glan-Taylor (GT) prism forming a slightly converging p-polarized beam on the surface of the sample. The intensity of the specularly reflected beam (zero diffraction order) was registered by an Ocean Optics Flame spectrometer. Two magnets with opposite magnetization directions were used to generate the static external magnetic fields with a strength of 650~G parallel to the nanowire axis for TMOKE measurements. The TMOKE value $\delta$ was calculated according to the following equation:
\begin{equation}
\label{eq:0}
\delta =2 \times \frac{R(\bm{M})-R(\bm{-M})}{R(\bm{M})+R(\bm{-M})} \times 100\%,
\end{equation}
where $R(\bm{M})$ and $R(\bm{-M})$ are the reflectances corresponding to opposite magnetization directions.

The numerical simulations of reflectance and TMOKE spectra, as well as the near-field distributions, were carried out using the FDTD (finite-difference time-domain) method in the Ansys Lumerical FDTD software~\cite{lumerical1}. The light source was modeled as a plane wave. The permittivity tensor of nickel was taken in the form that depends on the magnetization directed either along or against \textit{z} axis:
\begin{equation} \label{eq:1}
\hat{\varepsilon} = 
\begin{pmatrix} 
\varepsilon_1 & \pm ig & 0\\
\mp ig & \varepsilon_1 & 0 \\
0 & 0 & \varepsilon_1\\
\end{pmatrix},
\end{equation}
where $\varepsilon_{1}$ is the dielectric permittivity and $g = |\bm{g}|$ is the gyration constant. The permittivity tensor values of silicon and nickel were taken from the experimental data in Ref.~\cite{alpha-si, palik, Krinchik1968MagnetoopticalPO}. 
\section{Results and discussion}

Figure~~\ref{fig:1}~(a1 -- a4) presents the schematic cross-sections of the structures and the geometrical parameters used in the simulations. The blue curves in Figure~\ref{fig:1}~(b1 -- b4, first row) show the experimental reflectance spectra in the spectral range from $550$~nm to $900$~nm at optimal (in terms of TMOKE enhancement) angles of incidence for samples with varying width (see the relevant discussion below about angular-wavelength TMOKE dependences). The dips in these spectra are caused by the Mie mode excitation in the nanowires. To understand the nature of these resonances we carried out the numerical simulations of the reflectance spectra, presented in Figure~\ref{fig:1}~(b1 -- b4, second row), and the optical magnetic $|\textit{\textbf{H}}(x,y)|$ near-field distributions across the silicon nanowire and nickel substrate depicted in Figure~\ref{fig:1}~(c1 -- c4) at the resonant wavelengths. A good agreement with the experimental data is achieved. 

\begin{figure*}[!b]
    \centering
    \includegraphics[width=0.9\linewidth]{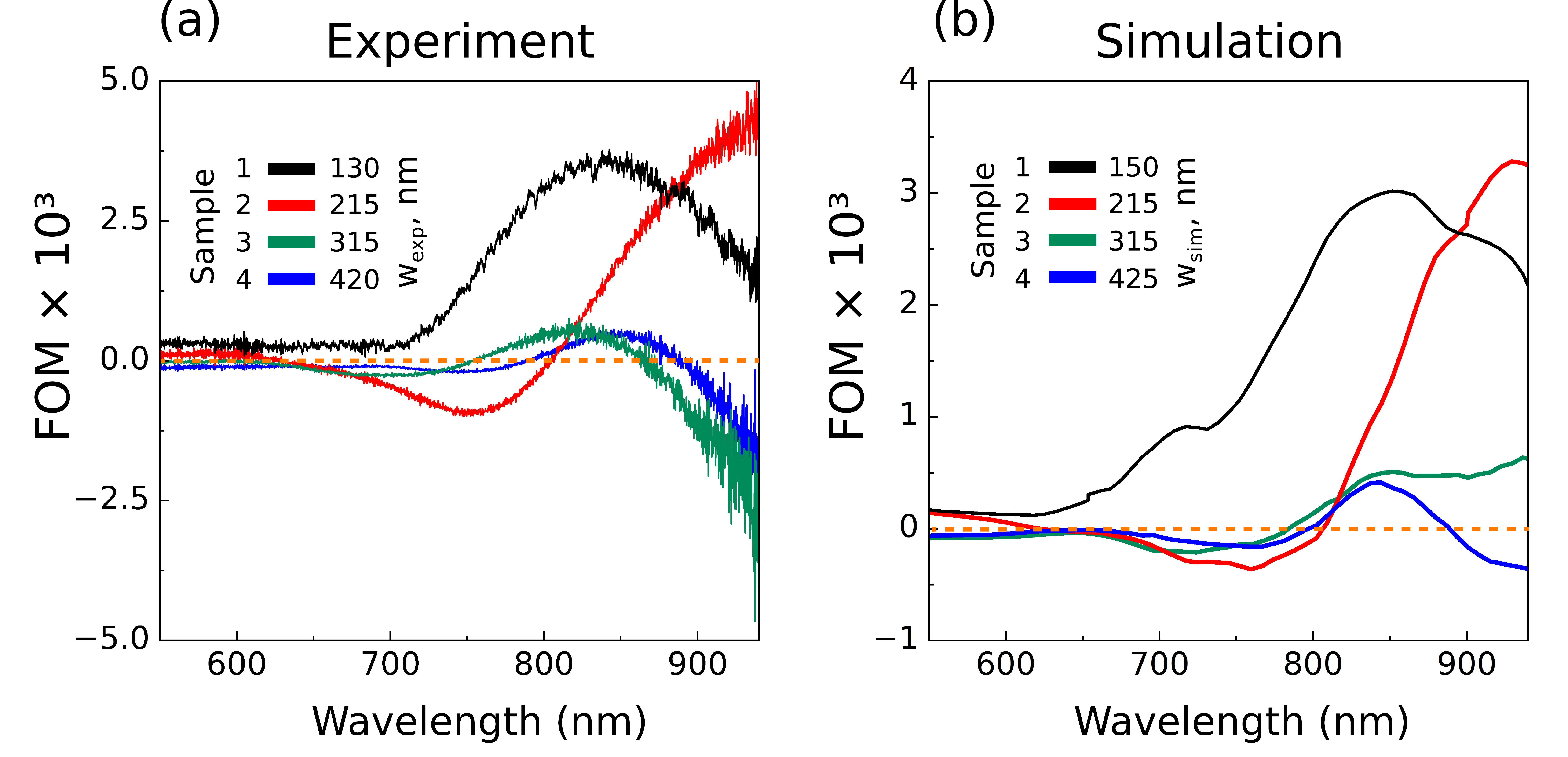}
	\caption{Experimental (a) and simulated (b) FOM ($\Delta R$) spectra for Si-Ni metasurfaces (Samples 1 –- 4) with different nanowire widths.}
	\label{fig:2}
\end{figure*}

The spectrally broad dip in the reflectance spectrum of Sample 1 with a minimal nanowire width at $\lambda=765$~nm (in simulations) (Figure~\ref{fig:1}~(b1), blue curves) is caused by the excitation of the magnetic dipole mode (MD$_z$) oriented along the nanowire \textit{z} axis (see a single antinode in the magnetic field in Figure~\ref{fig:1}~(c1)). An increase in the nanowire width led to a spectral redshift of MD$_z$ resonances excited in Samples 2 and 3 at $\lambda=840$~nm and $\lambda=820$~nm respectively, see Figure~\ref{fig:1}~(c2,~c3). Besides, Samples 2, 3 and 4 also support the excitation of the electric modes in the considered wavelength range: electric dipole ED$_y$ inclined with respect to \textit{y} axis at $\lambda=725$~nm (Figure~\ref{fig:1}~(c2)), electric quadrupole EQ at $\lambda=710$~nm and $\lambda=750$~nm respectively, see Figure~\ref{fig:1}~(c3,~c4). 

The optical magnetic and electric near-fields of the excited Mie modes are mostly localized inside the volume of the nanowire. However, they induce electric current oscillations in the nickel layer, leading to the strong field localization at the Si/Ni interface as shown in Figure~\ref{fig:1}~(c1~-~c4). Under external transverse magnetization, the anisotropic dielectric permittivity tensor (Eq.~\ref{eq:1}) allows for coupling of $E_x$ and $E_y$ components that further leads to a strong change in reflectance, i.e., the TMOKE. The maxima of the TMOKE (red curves in Figure~\ref{fig:1}~(b1 -- b4)) lie in the spectral vicinity of the reflectance minima associated with the excitation of Mie modes. The reconfiguration of the near-field distribution at the Si/Ni interface by changing the silicon nanowire width tunes the maximal TMOKE value. Sample 1 possesses its largest experimental TMOKE value of $\delta = 2.2\%$ in the vicinity of MD$_z$ mode excitation. Note that the reference TMOKE values in the considered spectral range for a nickel film~\cite{Krinchik1968MagnetoopticalPO} and in the non-resonant spectral ranges for Samples 1~--~4 do not exceed $\delta~=~0.5\%$.
However, the field localization at the Si/Ni interface is weak for a small nanowire width. Increasing the nanowire width enhanced the electromagnetic field at the Si/Ni interface under MD$_z$ excitation, and the maximum TMOKE value reached 3.8\% for Samples 2 and 3 (see Figure~\ref{fig:1}~(b2, b3)). 
The excitation of other modes resulted in a weaker TMOKE enhancement, see Figure~\ref{fig:1}~(b3, b4). The simulated TMOKE spectra in Figure~\ref{fig:1}~(b1 -- b4, second row) support these observations. The largest TMOKE values were thus caused by the MD$_z$ mode excitation in Samples 2 and 3, both in the experiment and the simulations. The measured TMOKE values can be tailored from $\delta = 2.2\% $ to $\delta = 3.8\%$ by changing the near-field distribution of the MD$_z$ mode through the variation in the nanowire width. 
Moreover, the spectral bandwidth of TMOKE resonant enhancement can be tuned. For example, wideband effect enhancement due to the individual MD$_z$ mode was observed in Sample~1 in the spectral range from $\lambda = 700$~nm to~$\lambda = 900$~nm, while a more narrowband enhancement occurred in Sample~3 in the vicinity of $\lambda = 800$~nm in the experiment. 
\begin{figure*}[!b]
	\centering
	\includegraphics[width=0.95\linewidth]{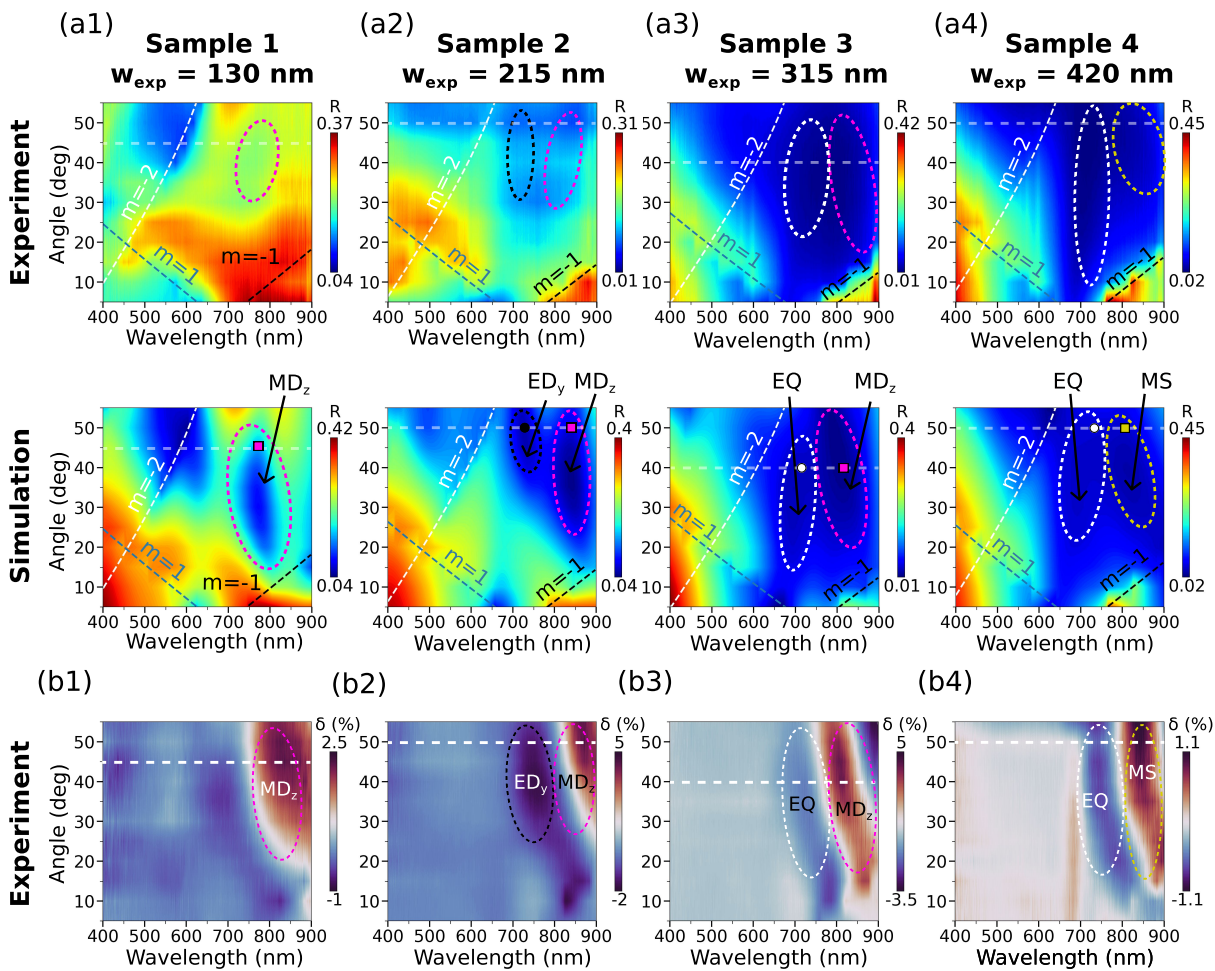}
	\caption{(a1 -- a4) Experimental (first row) and simulated (second row) angular-wavelength reflectance dependences for Si-Ni metasurfaces (Samples 1 -- 4). (b1 -- b4) Experimental TMOKE ($\delta$) angular-wavelength dependences for Si-Ni metasurfaces (Samples 1 -- 4). The horizontal dashed lines on all maps highlight the angles of incidence at which the maximum TMOKE value was achieved. The oval dashed curves indicate the area of Mie modes excitation and the magneto-optical enhancement associated with these modes in TMOKE maps.}
	\label{fig:3}
\end{figure*} 

The nanowire width affects the distributions of \textit{E$_{x}$} and \textit{E$_{y}$} field components in the nickel layer, which are linked to both $R$ and the change in reflectance $\Delta R = R(\bm{M})-R(\bm{-M})$ (referred to as the figure of merit) occurring in the external magnetic field~\cite{MO_theory}. The non-resonant reflection increases if the nanowire width is decreased (one can see the trend in reflectance spectra in Figure~\ref{fig:1}). The resulting reflectance can be described as a Fano-type interference between non-resonant reflection and Mie resonant scattering~\cite{Fano, Limonov2017}. Thus, such a trade-off between $\Delta R$ and $R$ leads to an optimal configuration for TMOKE ($\frac{\Delta R}{R}$). Besides, this observation explains the fact that $\frac{\Delta R}{R}$ and $\Delta R$ maxima occur at different nanowire widths, as shown by the experimental and simulated FOM spectra at the corresponding optimal angles of light incidence demonstrated in Figure~\ref{fig:2}~(a) and (b). The largest FOM values were observed in Sample 2 in the spectral vicinity of the MD$_z$ mode excitation, while the largest TMOKE was experimentally observed in Samples~2 and 3. A much smaller FOM for Sample~3 than Sample~2 means that the TMOKE was enhanced due to small reflectance values. It has a negative impact on the sensing applications~\cite{10.1063/5.0072884} due to a decrease in the signal-to-noise ratio. Figure~\ref{fig:2} shows that the Si-Ni metasurfaces also allow for a significant FOM tunability. By changing the nanowire width FOM can be varied from $\Delta R = 0.5 \times 10^{-3}$ (Sample~4) to $\Delta R = 4.4 \times 10^{-3}$ (Sample~2) in the range of $ \lambda = 800$ -- $900$~nm in the experiment (similarly from $\Delta R = 0.4~\times~10^{-3}$ to $\Delta R = 3.3\times 10^{-3}$ in the simulations). Compared to the previous works on magneto-optics in Si-Ni nanostructures, not only did we experimentally observe a significant tunability of the TMOKE, but we also reached larger FOM and TMOKE values. For instance, the previously discussed metasurfaces, which consist of a silicon nanodisk rectangular array covered with a 5 nm Ni film in Ref.~\cite{nanolabMiemagnetic} demonstrate $\delta = 0.5\%$ and $\Delta T = 2 \times 10^{-3}$ (FOM measured in transmittance geometry), while in this paper we demonstrate up to $\delta = 3.8\%$ and $\Delta R = 4.4 \times 10^{-3}$. 

The aforementioned reflectance and TMOKE spectra were presented only for the optimal angles of incidence, and the full angular dependence of these quantities should be discussed. The top row in Figure~\ref{fig:3}~(a1 -- a4) shows the experimental reflectance spectra in the spectral range from $\lambda = 400$~nm to $\lambda = 900$~nm for the angles of light incidence from 5$^{\circ}$ to 55$^{\circ}$ which were measured with a step of 5$^{\circ}$. The bottom row in the same figure shows the corresponding simulated reflectance dependences. The areas of reflectance minima and maxima can be observed on both the experimental and simulated maps. The metasurfaces under study represent one-dimensional nanogratings, and, therefore, various diffraction orders are observed. The angle-dependent maxima of reflectance correspond to Rayleigh anomalies (RAs), which occur when one of the diffraction orders propagates along the surface of the grating. The spectral-angular position of RAs was defined by the following equations derived directly from the diffraction grating equation: $sin(\varphi_{-1}) = -1+\lambda/d$, $sin(\varphi_{-2}) = -1+2\lambda/d$, $sin(\varphi_{+1}) = 1-\lambda/d$ for RAs corresponding to the $-1^{st}$, $-2^{d}$, and $+1^{st}$ diffraction orders (m), respectively. The straight dashed lines which demonstrate the RA dispersion are superimposed on the experimental and simulated reflectance colormaps in Figure~\ref{fig:3}~(a1 -- a4). In contrast to the sharp angular dependent maxima associated with RAs, there are angular-independent reflectance minima in the spectral range from 600~nm to 900~nm and incident angles from $15^{\circ}$ to $50^{\circ}$ marked by oval colored dashed curves. These minima were caused by the excitation of the Mie modes localized in silicon nanowires. The oval pink dashed curves in Figure~\ref{fig:3}~(a1 -- a3) show the spectral-angular areas of the MD$_z$ mode excitation. The areas of other excited modes (ED$_y$, EQ, MS) are also depicted on the maps by oval dashed curves. The corresponding |$\textit{\textbf{H}}(x,y)$| field distributions at the spectral-angular positions marked with colored symbols on the reflectance maps (pink square for the MD$_z$ mode, etc.) are shown previously in Figure~\ref{fig:1}~(c1 -- c4). 

Figure~\ref{fig:3}~(b1 -- b4) presents the experimental angular-wavelength TMOKE dependences, which reveal the influence of Mie modes and RAs. Oval dashed curves highlight the areas of maximum positive and negative $\delta$ values representing resonant TMOKE enhancement. The spectral position of $\delta$ maxima in these areas weakly depends on the angle of light incidence, and these maxima emerge due to the excitation of the corresponding Mie modes. The white dashed horizontal lines highlight the angles at which the TMOKE reached the largest values (the reflectance spectra and field distributions in Figure~\ref{fig:1} are presented for these optimal angles of incidence). The TMOKE values near RAs are similar to the non-resonant ones and do not show strong angular-wavelength dependence that is typical of RAs. Thus, the contribution of RAs and the possible excitation of surface plasmons to the TMOKE was negligible compared to that of Mie resonances. 

To~conclude, we have proposed a hybrid metal-dielectric metasurface consisting of rectangular silicon nanowires on a nickel substrate, which enables the control of the TMOKE and FOM values by changing the width of the nanowires. At certain nanowire widths, the in-plane magnetic dipole mode excitation leads to both near-field localization and optimal far-field response that provide maximum TMOKE and FOM values up to $\delta = 3.8\%$ and $\Delta R = 4.4 \times 10^{-3}$, respectively. Besides, TMOKE and FOM enhancement spectral bandwidth and position can also be reconfigured, opening up possibilities for new applications in integrated~\cite{Kristian, Salazar2023, Salazar2022_1} and ultrafast magnetooptics~\cite{Novikov2020, Novikov2023, Temnov2023_2, Temnov2023_1, Chekhov-21}.

\medskip
\textbf{Acknowledgements.} 
The authors have no conflicts of interest to declare. The work is funded by the Russian Science Foundation Grant No. 24-72-00042. This research was performed according to the Development Program of the Interdisciplinary Scientific and Educational School of Lomonosov Moscow State University “Photonic and Quantum Technologies. Digital Medicine.”
\medskip

\bibliographystyle{MSP}
{\small
\bibliography{bibliography}}
\end{document}


\title{Supporting Information: Tailoring Transverse Magneto-Optical Kerr Effect Enhancement in Mie-resonant Nanowire-based Metasurfaces}
\author{Karen A. Mamian} \date{}
\author{Vladimir V. Popov} \date{}
\author{Aleksandr Yu. Frolov} \date{}
\author{Andrey A. Fedyanin} \date{}
\affil{Faculty of Physics, Lomonosov Moscow State University, Moscow 119991, Russia}

\maketitle
\section*{Atomic force microscopy (AFM) data}
\renewcommand{\figurename}{}
\begin{figure}[h!]
	\centering
	\includegraphics[width=0.8\linewidth]{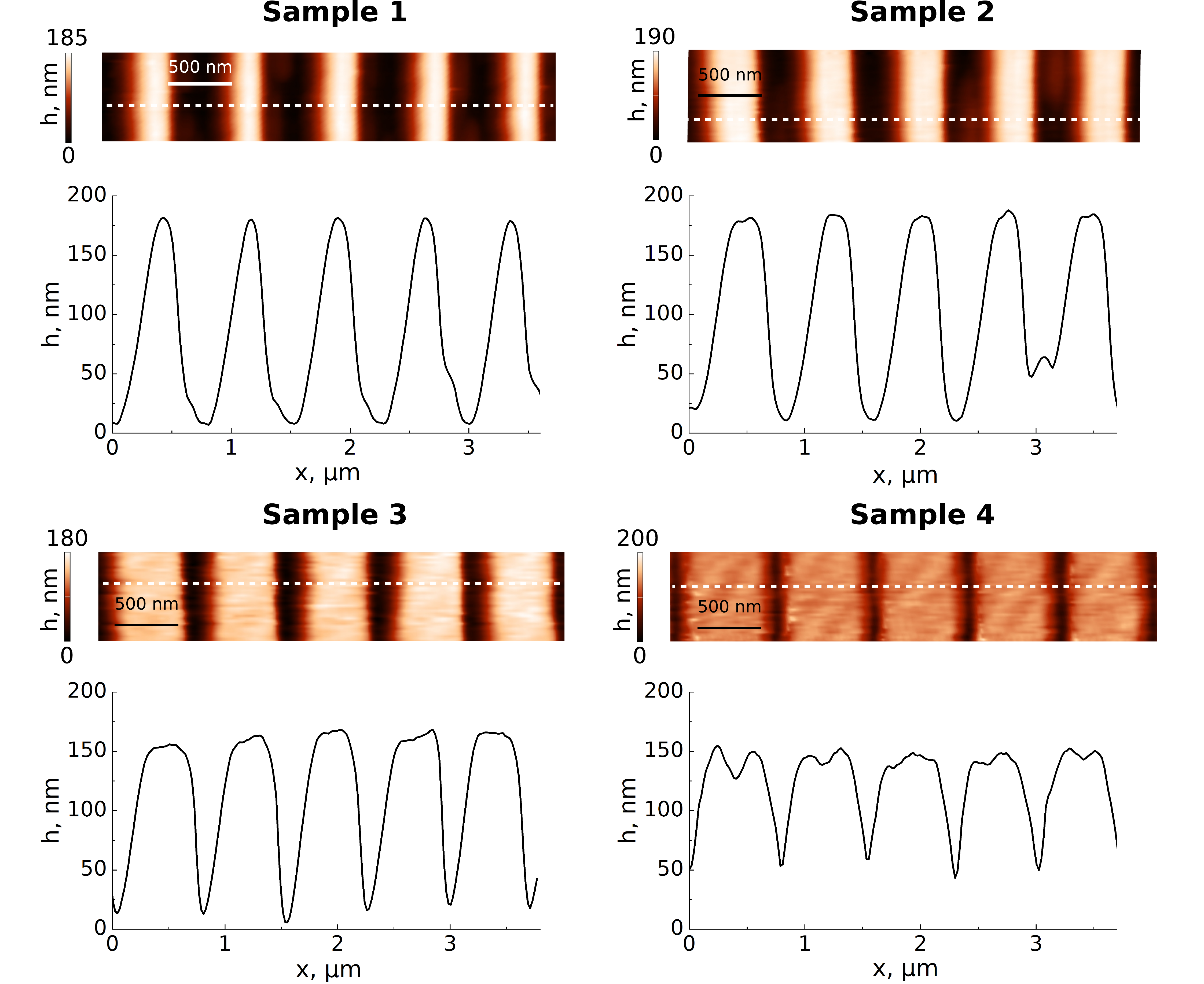}
	\caption*{Figure S1: AFM images and their cross-sections along the dashed lines for all samples.}
	\label{fig:0}
\end{figure}

The nanowire heights were measured using the AFM data presented in Figure~S1. Samples 1 and 2 are 170 -- 180~nm high, which corresponds to the heights used in numerical simulations. However, the AFM images do not provide a valid height value for Samples 3 and 4 since the nanowires are too close to each other for the AFM tip to properly scan the surface. Nevertheless, the samples were fabricated from the same silicon film during one fabrication process, so Samples 3 and 4 were assumed to have a height of 170~nm in the models, and numerical simulations prove the accuracy of this assumption.